# SARS-CoV-2 and miRNA-like inhibition power

**Jacques Demongeot**[1*], **Hervé Seligmann**[1,2]

[1] Laboratory AGEIS EA 7407, Team Tools for e-Gnosis Medical & Labcom CNRS/UGA/OrangeLabs Telecom4Health, Faculty of Medicine, University Grenoble Alpes (UGA), 38700 La Tronche, France; Jacques.Demongeot@univ-grenoble-alpes.fr

[2] The National Natural History Collections, The Hebrew University of Jerusalem, 91404 Jerusalem, Israel; varanuseremius@gmail.com

\* Correspondence: Jacques.Demongeot@univ-grenoble-alpes.fr

**Abstract:** (1) Background: RNA viruses and especially coronaviruses could act inside host cells not only by building their own proteins, but also by perturbing the cell metabolism. We show the possibility of miRNA-like inhibitions by the SARS-CoV-2 concerning for example the hemoglobin and type I interferons syntheses, hence highly perturbing oxygen distribution in vital organs and immune response as described by clinicians; (2) Methods: We compare RNA subsequences of SARS-CoV-2 protein S and RNA-dependent RNA polymerase genes to mRNA sequences of beta-globin and type I interferons; (3) Results: RNA subsequences longer than eight nucleotides from SARS-CoV-2 genome could hybridize subsequences of the mRNA of beta-globin and of type I interferons; (4) Conclusions: Beyond viral protein production, Covid-19 might affect vital processes like host oxygen transport and immune response.

## 1. Introduction

Viruses act in host cells by reproducing their own proteins for reconstituting their capsid, duplicating their genome [1] and leaving non-coding RNA or DNA remnants in host genomes [2]. Moreover, RNA viruses can also form complexes with existing mRNAs and/or proteins of host cells. Thereby they might prevent protein function, behave like microRNAs [3-6] or ribosomal RNAs [6-8], inhibiting or favoring the translation of specific proteins of host cells [9-17]. If these proteins are vital for the host, viral pathogenicity is much greater than that caused by viral replication. With regard to SARS-CoV-2, binding to existing host proteins has already been described [18]. Here, we aim to describe a potential miRNA-like action by viral RNA, in particular i) at the level of oxygen transport by hemoglobin, whose beta-globin and gamma 2 subunits synthesis can be inhibited, and ii) at the level of immune response, where type I interferon synthesis can be inhibited. We are not intending to prove here experimentally these inhibitions by small RNAs issued from the SARS-CoV-2 genome, but to prepare this future empirical step by pointing out its potential hybridizing power. In Section 2, we describe a method for finding SARS-CoV-2 inhibitory RNA sub-sequences, and results are given in Section 3, ~~we~~ discussed in Section 4. Some perspectives of this work concerning an extension to the inhibition of translation of olfactory and interferon receptors are proposed in Section 5.

## 2. Methods

Focusing on the seed part of miRNA-like sequences having a putative 8 nucleotide hybridization seed inhibition effect [19-20] (minimum 7), we compare data from different databases

[21-26] using BLAST [27]. Figure 1 shows microRNA 129-5p, a known inhibitor of a human foetal hemoglobin component, the gamma-globin 2, replaced in adult by the beta-globin regulated as the other component alpha-globin, by microRNAS [28-32]. Two sub-sequences from the SARS-CoV-2 genome, namely from genes of ORF10 and protein S, show the same hybridizing potential.

**Homo sapiens hemoglobin subunit gamma 2 (HBG2), mRNA NCBI Reference Sequence: NM_000184.3**

```
        5'–ACACTCGCTTCTGGAACGTCTGAGGTTATCAATAAGCTCCTAGTCCAGACGCCATGGGTCATTTCACAGA
    GGAGGACAAGGCTACTATCACAAGCCTGTGGGCAAGGTGAATGTGGAAGATGCTGGAGGAGAAACCCTG
    GGAAGGCTCCTGGTTGTCTACCCATGGACCCAGAGGTTCTTTGACAGCTTTGGCAACCTGTCCTCTGCCT
    CTGCCATCATGGGCAACCCCAAAGTCAAGGCACATGGCAAGAAGGTGCTGACTTCCTTGGGAGATGCCAT
                  hsa-miR-129-5p  3'-TGTCGTTCGGGTCTGGC-5'
    AAAGCACCTGGATGATCTCAAGGGCACCTTTGCCCAGCTGAGTGAACTGCACTGTGACAAGCTGCATGTG
    GATCCTGAGAACTTCAAGCTCCTGGGAAATGTGCTGGTGACCGTTTTGGCAATCCATTTCGGCAAAGAAT
    TCACCCCTGAGGTGCAGGCTTCCTGGCAGAAGATGGTGACTGGAGTGGCCAGTGCCCTGTCCTCCAGATA
              protein S gene SARS-CoV-2  3'-CTGAAGTAGTGGAGATTAATGT-5'
    CCACTGAGCTCACTGCCCATGATGCAGAGCTTTCAAGGATAGGCTTTATTCTGCAAGCAATCAAATAATA
                 ORF10 gene SARS-CoV-2  3'-TTCTAAGTAAGACGTGTTCTCATCTG-5'
    AATCTATTCTGCTAAGAGATCACACA-3'
```

**Figure 1.** Complete mRNA sequence of the subunit gamma 2 of the fetal human hemoglobin [22]. Sequences in green (resp. red) come from protein S and ORF10 genes of SARS-CoV-2 (resp. hsa miR 129-5p), which can inhibit its ribosomal translation. Probability of length 8 anti-match in red (resp. 9 and 11 in green) by chance in 577 nucleotides equals 0.035 (resp. 0.017 and 0.0003) (T-G and G-T matches counting for ½).

## 3. Results

We will apply the method from Section 2 for showing examples where RNA subsequences of the SARS-CoV-2 genome have an inhibitory potential on the ribosomal translation of human mRNAs of the same type as that shown in Section 2 for human micro-RNAs. For example, miRTarBase shows that microRNA hsa-mir-92a-3p targets the beta-globin HBB subunit of adult hemoglobin, inhibiting its translation [25]. This is also the case for microRNAs involved in the maturation of erythrocytes like miR-451a [26-31]. We exhibit on Figure 2 subsequences of the SARS-CoV-2 protein S and polymerase genes [23] having the same length of antimatching as these microRNAs on the mRNA of the hemoglobin beta-globin (HBB) subunit gene.

**Homo sapiens hemoglobin subunit beta (HBB), mRNA NCBI Reference Sequence: NM_000518.5**

```
        5'–ACATTTGCTTCTGACACAACTGTGTTCACTAGCAACCTCAAACAGACACCATGGTGCATCTGACTCCTGAGGA
           RNA-dependent RNA polymerase SARS-CoV-2   3'-TCACGTAGAACTAGGAGTATT-5'
    GAAGTCTGCCGTTACTGCCCTGTGGGGCAAGGTGAACGTGGATGAAGTTGGTGGTGAGGCCCTGGGCAGGCTG
    CTGGTGGTCTACCCTTGGACCCAGAGGTTCTTTGAGTCCTTTGGGGATCTGTCCACTCCTGATGCTGTTATGG
    GCAACCCTAAGGTGAAGGCTCATGGCAAGAAAGTGCTCGGTGCCTTTAGTGATGGCCTGGCTCACCTGGACA
                      mir-451a   3'-TTGAGTCATTACCATTGCCAA-5'
    CCTCAAGGGCACCTTTGCCACACTGAGTGAGCTGCACTGTGACAAGCTGCACGTGGATCCTGAGAACTTCAGG
                                                                   3'-TGTCCGGC
    CTCCTGGGCAACGTGCTGGTCTGTGTGCTGGCCCATCACTTTGGCAAAGAATTCACCCCACCAGTGCAGGCTG
    CCTGTTCACGTTAT-5' miR-92a-3p
    CCTATCAGAAAGTGGTGGCTGGTGTGGCTAATGCCCTGGCCCACAAGTATCACTAAGCTCGCTTTCTTGCTGT
       3'-ATTTTCACCTTTTACTACGCC-5' Protein S SARS-CoV-2
    CCAATTTCTATTAAAGGTTCCTTTGTTCCCTAAGTCCAACTACTAAACTGGGGGATATTATGAAGGGCCTTGA
    GCATCTGGATTCTGCCTAATAAAAAACATTTATTTTCATTGCAA-3'
```

**Figure 2.** Human beta-globin gene [24] potentially targeted by a subsequence of the gene of the SARS-CoV-2 RNA-dependent RNA polymerase (in blue) and by a subsequence of the gene of the SARS-CoV-2 protein S (in green) [23], by the human microRNAs hsa miR 92a-3p (in red) and hsa miR 451a (in red). Probability of red anti-matches of length 8 in a sequence of 624 nucleotides equals 0.04 and for the blue (resp. green) subsequence is 0.005 (resp. 0.017) (T-G and G-T matches counting for ½).

The second example concerns the gene of the spicule protein S of SARS-CoV-2, which shares a long subsequence of length 14 (664-678) with the gene of the Gag protein of the virus HERV-K102 (Figure 3). Its potential targets are the mRNAs of human hemoglobin subunit beta-globin [22], human hemoglobin subunit gamma-globin 2 (HBG2) [23], human type 1 interferons and the human receptor ACE2.

**SARS coronavirus 2 isolate USA/MN1-MDH1/2020, complete genome GenBank: MT188341.1: 21512-25333 protein S**

```
5'-ATGTTTGTTTTTCTTGTTTTATTGCCACTAGTCTCTAGTCAGTGTGTTAATCTTACAACCAGAACTCAAT
TACCCCCTGCATACACTAATTCTTTCACACGTGGTGTTTATTACCCTGACAAAGTTTTCAGATCCTCAGT
TTTACATTCAACTCAGGACTTGTTCTTACCTTTCTTTTCCAATGTTACTTGGTTCCATGCTATACATGTC
TCTGGGACCAATGGTACTAAGAGGTTTGATAACCCTGTCCTACCATTTAATGATGGTGTTTATTTTGCTT
CCACTGAGAAGTCTAACATAATAAGAGGCTGGATTTTTGGTACTACTTTAGATTCGAAGACCCAGTCCCT
ACTTATTGTTAATAACGCTACTAATGTTGTTATTAAAGTCTGTGAATTTCAATTTTGTAATGATCCATTT
TTGGGTGTTTATTACCACAAAACAACAAAAGTTGGATGGAAAGTGAGTTCAGAGTTTATTCTAGTGCGA
ATAATTGCACTTTTGAATATGTCTCTCAGCCTTTTCTTATGGACCTTGAAGGAAAACAGGGTAATTTCAA
AAATCTTAGGGAATTTGTGTTTAAGAATATTGATGGTTATTTTAAAATATATTCTAAGCACACGCCTATT
AATTTAGTGCGTGATCTCCCTCAGGGTTTTTCGGCTTTAGAACCATTGGTAGATTTGCCAATAGGTATTA
ACATCACTAGGTTTCAAACTTTACTTGCTTTACATAGAAGTTATTTGACTCCTGGTGATTCTTCTTCAGG
TTGGACAGCTGGTGCTGCAGCTTATTATGTGGGTTATCTTCAACCTAGGACTTTTCTATTAAAATATAAT
GAAAATGGAACCATTACAGATGCTGTAGACTGTGCACTTGACCCTCTCTCAGAAACAAAGTGTACGTTGA
AATCCTTCACTGTAGAAAAAGGAATCTATCAAACTTCTAACTTTAGAGTCCAACCAACAGAATCTATTGT
TAGATTTCCTAATATTACAAACTTGTGCCCTTTTGGTGAAGTTTTTAACGCCACCAGATTTGCATCTGTT
TATGCTTGGAACAGGAAGAGAATCAGCAACTGTGTTGCTGATTATTCTGTCCTATATAATTCCGCATCAT
TTTCCACTTTTAAGTGTTATGGAGTGTCTCCTACTAAATTAAATGATCTCTGCTTTACTAATGTCTATGC
AGATTCATTTGTAATTAGAGGTGATGAAGTCAGACAAATCGCTCCAGGGCAAACTGGAAAGATTGCTGAT
TATAATTATAAATTACCAGATGATTTTACAGGCTGCGTTATAGCTTGGAATTCTAACAATCTTGATTCTA
AGGTTGGTGGTAATTATAATTACCTGTATAGATTGTTTAGGAAGTCTAATCTCAAACCTTTTGAGAGAGA
TATTTCAACTGAAATCTATCAGGCCGGTAGCACACCTTGTAATGGTGTTGAAGGTTTTAATTGTTACTTT
CCTTTACAATCATATGGTTTCCAACCCACTAATGGTGTTGGTTACCAACCATACAGAGTAGTAGTACTTT
CTTTTGAACTTCTACATGCACCAGCAACTGTTTGTGGACCTAAAAAGTCTACTAATTTGGTTAAAAACAA
ATGTGTCAATTTTAACTTCAATGGTTTAACAGGCACAGGTGTTCTTACTGAGTCTAACAAAAAGTTTCTG
CCTTTCCAACAATTTGGCAGAGACATTGCTGACACTACTGATGCTGTCCGTGATCCACAGACACTTGAGA
TTCTTGACATTACACCATGTTCTTTTGGTGGTGTCAGTGTTATAACACCAGGAACAAATACTTCTAACCA
GGTTGCTGTTCTTTATCAGGATGTTAACTGCACAGAAGTCCCTGTTGCTATTCATGCAGATCAACTTACT
CCTACTTGGCGTGTTTATTCTACAGGTTCTAATGTTTTTCAAACACGTGCAGGCTGTTTAATAGGGGCTG
AACATGTCAACAACTCATATGAGTGTGACATACCCATTGGTGCAGGTATATGCGCTAGTTATCAGACTCA
GACTAATTCTCCTCGGCGGGCACGTAGTGTAGCTAGTCAATCCATCATTGCCTACACTATGTCACTTGGT
GCAGAAAATTCAGTTGCTTACTCTAATAACTCTATTGCCATACCCACAAATTTTACTATTAGTGTTACCA
CAGAAATTCTACCAGTGTCTATGACCAAGACATCAGTAGATTGTACAATGTACATTTGTGGTGATTCAAC
TGAATGCAGCAATCTTTTGTTGCAATATGGCAGTTTTTGTACACAATTAAACCGTGCTTTAACTGGAATA
GCTGTTGAACAAGACAAAAACACCCAAGAAGTTTTTGCACAAGTCAAACAAATTTACAAAACACCACCAA
TTAAAGATTTTGGTGGTTTTAATTTTTCACAAATATTACCAGATCCATCAAAACCAAGCAAGAGGTCATT
TATTGAAGATCTACTTTTCAACAAAGTGACACTTGCAGATGCTGGCTTCATCAAACAATATGGTGATTGC
CTTGGTGATATTGCTGCTAGAGACCTCATTTGTGCACAAAAGTTTAACGGCCTTACTGTTTTGCCACCTT
TGCTCACAGATGAAATGATTGCTCAATACACTTCTGCACTGTTAGCGGGTACAATCACTTCTGGTTGGAC
CTTTGGTGCAGGTGCTGCATTACAAATACCATTTGCTATGCAAATGGCTTATAGGTTTAATGGTATTGGA
GTTACACAGAATGTTCTCTATGAGAACCAAAAATTGATTGCCAACCAATTTAATAGTGCTATTGGCAAAA
TTCAAGACTCACTTTCTTCCACAGCAAGTGCACTTGGAAAACTTCAAGATGTGGTCAACCAAAATGCACA
AGCTTTAAACACGCTTGTTAAACAACTTAGCTCCAATTTTGGTGCAATTTCAAGTGTTTTAAATGATATC
CTTTCACGTCTTGACAAAGTTGAGGCTGAAGTGCAAATTGATAGGTTGATCACAGGCAGACTTCAAAGTT
TGCAGACATATGTGACTCAACAATTAATTAGAGCTGCAGAAATCAGAGCTTCTGCTAATCTTGCTGCTAC
TAAAATGTCAGAGTGTGTACTTGGACAATCAAAAAGAGTTGATTTTTGTGGAAAGGGCTATCATCTTATG
TCCTTCCCTCAGTCAGCACCTCATGGTGTAGTCTTCTTGCATGTGACTTATGTCCCTGCACAAGAAAGA
ACTTCACAACTGCTCCTGCCATTTGTCATGATGGAAAAGCACACTTTCCTCGTGAAGGTGTCTTTGTTTC
AAATGGCACACACTGGTTTGTAACACAAAGGAATTTTTATGAACCACAAATCATTACTACAGACAACACA
TTTGTGTCTGGTAACTGTGATGTTGTAATAGGAATTGTCAACAACACAGTTTATGATCCTTTGCAACCTG
AATTAGACTCATTCAAGGAGGAGTTAGATAAATATTTTAAGAATCATACATCACCAGATGTTGATTTAGG
TGACATCTCTGGCATTAATGCTTCAGTTGTAAACATTCAAAAAGAAATTGACCGCCTCAATGAGGTTGCC
AAGAATTTAAATGAATCTCTCATCGATCTCCAAGAACTTGGAAAGTATGAGCAGTATATAAAATGGCCAT
GGTACATTTGGCTAGGTTTTATAGCTGGCTTGATTGCCATAGTAATGGTGACAATTATGCTTTGCTGTAT
GACCAGTTGCTGTAGTTGTCTCAAGGGCTGTTGTTCTTGTGGATCCTGCTGCAAATTTGATGAAGACGAC
TCTGAGCCAGTGCTCAAAGGAGTCAAATTACATTACACATAA-3'
```

**Figure 3.** mRNA sequence of the protein S of the virus SARS-CoV-2 [23]. The first green subsequence of length 14 (664-678) occurs in mRNA of Gag protein of the virus HERV-K102 [27]. The second of length 23 (1112-1134) anti-matches a mRNA subsequence of hemoglobin subunit beta-globin [22]. The third of length 22 (1200-1221) anti-matches a mRNA subsequence of hemoglobin subunit gamma-globin 2 (HBG2) [23]. The fourth of length 24 (2032-2055) matches a subsequence of mRNA of many type 1 interferons. Highlighted in yellow are sub-sequences common with the SARS furin cleavage site [33-34]. The fifth of length 25 (3152-3176) matches a subsequence of mRNA of the receptor ACE2. Blue: mutations whose location of both codon and nucleotide involved [35] are, in order: 635 gCtagTt, 1133 aAgaaGg, 2045 cGgacAg and 3189 ttG>ttT. The probabilities of the above matches and anti-matches will be given in the following in Figures concerning each of them.

**Homo sapiens ACE2 mRNA, complete cds GenBank: AB046569.1**

```
5'-TTTTTAGTCTAGGGAAAGTCATTCAGTGGATGTGATCTTGGCTCACAGGGGACGATGTCAAGCTCTTCCT
GGCTCCTTCTCAGCCTTGTTGCTGTAACTGCTGCTCAGTCCACCATTGAGGAACAGGCCAAGACATTTTT
GGACAAGTTTAACCACGAAGCCGAAGACCTGTTCTATCAAAGTTCACTTGCTTCTTGGAATTATAACACC
                  protein S SARS-CoV-2      3'-CTGTTTACCGTCCTCGTCAACACTT-5'
AATATTACTGAAGAGAATGTCCAAAACATGAATAACGCTGGGGACAAATGGTCTGCCTTTTTAAAGGAAC
AGTCCACACTTGCCCAAATGTATCCACTACAAGAAATTCAGAATCTCACAGTCAAGCTTCAGCTGCAGGC
TCTTCAGCAAAATGGGTCTTCAGTGCTCTCAGAAGACAAGAGCAAACGGTTGAACACAATTCTAAATACA
ATGAGCACCATCTACAGTACTGGAAAAGTTTGTAACCCAGATAATCCACAAGAATGCTTATTACTTGAAC
CAGGTTTGAATGAAATAATGGCAAACAGTTTAGACTACAATGAGAGGCTCTGGGCTTGGGAAAGCTGGAG
ATCTGAGGTCGGCAAGCAGCTGAGGCCATTATATGAAGAGTATGTGGTCTTGAAAAATGAGATGGCAAGA
GCAAATCATTATGAGGACTATGGGATTATTGGAGAGGAGACTATGAAGTAAATGGGGTAGATGGCTATG
ACTACAGCCGCGGCCAGTTGATTGAAGATGTGGAACATACCTTTGAAGAGATTAAACCATTATATGAACA
TCTTCATGCCTATGTGAGGGCAAAGTTGATGAATGCCTATCCTTCCTATATCAGTCCAATTGGATGCCTC
CCTGCTCATTTGCTTGGTGATATGTGGGGTAGATTTTGGACAAATCTGTACTCTTTGACAGTTCCCTTTG
GACAGAAACCAAACATAGATGTTACTGATGCAATGGTGGACCAGGCCTGGGATGCACAGAGAATATTCAA
GGAGGCCGAGAAGTTCTTTGTATCTGTTGGTCTTCCTAATATGACTCAAGGATTCTGGGAAAATTCCATG
CTAACGGACCCAGGAAATGTTCAGAAAGCAGTCTGCCATCCCACAGCTTGGGACCTGGGGAAAGGCGACT
TCAGGATCCTTATGTGCACAAAGGTGACAATGGACGACTTCCTGACAGCTCATCATGAGATGGGGCATAT
TCAGTATGATATGGCATATGCTGCACAACCTTTTCTGCTAAGAAATGGAGCTAATGAAGGATTCCATGAA
GCTGTTGGGAAATCATGTCACTTTCTGCAGCCACACCTAAGCATTTAAAATCCATTGGTCTTCTGTCAC
CCGATTTTCAAGAAGACAATGAAACAGAAATAAACTTCCTGCTCAAACAAGCACTCACGATTGTTGGGAC
TCTGCCATTTACTTACATGTTAGAGAAGTGGAGGTGGATGGTCTTTAAAGGGGAAATTCCCAAAGACCAG
TGGATGAAAAGTGGTGGGAGATGAAGCGAGAGATAGTTGGGGTGGTGGAACCTGTGCCCCATGATGAAA
CATACTGTGACCCCGCATCTCTGTTCCATGTTTCTAATGATTACTCATTCATTCGATATTACACAAGGAC
CCTTTACCAATTCCAGTTTCAAGAAGCACTTTGTCAAGCAGCTAAACATGAAGGCCCTCTGCACAAATGT
GACATCTCAAACTCTACAGAAGCTGGACAGAAACTGTTCAATATGCTGAGGCTTGGAAAATCAGAACCCT
GGACCCTAGCATTGGAAAATGTTGTAGGAGCAAAGAACATGAATGTAAGGCCACTGCTCAACTACTTTGA
GCCCTTATTTACCTGGCTGAAAGACCAGAACAAGAATTCTTTTGTGGGATGGAGTACCGACTGGAGTCCA
TATGCAGACCAAAGCATCAAAGTGAGGATAAGCCTAAAATCAGCTCTTGGAGATAGAGCATATGAATGGA
ACGACAATGAAATGTACCTGTTCCGATCATCTGTTGCATATGCTATGAGGCAGTACTTTTTAAAGTAAA
AAATCAGATGATTCTTTTTGGGGAGGAGGATGTGCGAGTGGCTAATTTGAAACCAAGAATCTCCTTTAAT
TTCTTTGTCACTGCACCTAAAAATGTGTCTGATATCATTCCTAGAACTGAAGTTGAAAAGGCCATCAGGA
TGTCCCGGAGCCGTATCAATGATGCTTTCCGTCTGAATGACAACAGCCTAGAGTTTCTGGGGATACAGCC
AACACTTGGACCTCCTAACCAGCCCCCTGTTTCCATATGGCTGATTGTTTTGGAGTTGTGATGGGAGTG
ATAGTGGTTGGCATTGTCATCCTGATCTTCACTGGGATCAGAGATCGGAAGAAGAAAAATAAAGCAAGAA
GTGGAGAAAATCCTTATGCCTCCATCGATATTAGCAAAGGAGAAAATAATCCAGGATTCCAAAACACTGA
TGATGTTCAGACCTCCTTTTAGAAAAATCTATGTTTTTCCTCTTGAGGTGATTTTGTTGTATGTAAATGT
TAATTTCATGGTATAGAAAATATAAGATGATAAAAATATCATTAAATGTCAAAACTATGACTCTGTTCAG-3'
```

**Figure 4.** mRNA sequence of the human protein receptor ACE2. The green 5'-3' seed subsequence of length 10 is the reverse of an RNA sequence of the protein S of SARS-CoV-2. The probability to observe such an anti-match of length 10 by chance in a sequence of 2581 nucleotides equals 0.003.

The classical protein-protein interaction of the spicule protein S of SARS-CoV-2 is with the human protein receptor ACE2, but there exists a putative miRNA-like translation inhibition due to a subsequence (in green) of the protein S gene (Figure 3) matching the ACE2 mRNA (Figure 4). The human endogenous retrovirus HERV-K102 [32] has been described as having an antagonizing power on HIV-1 replication, by stimulating antibody production. It is indeed capable of high replication rate *in vivo* and *in vitro* and this high particle production can stimulate an early protective innate immune response against HIV-1 replication. It could play the same role in SARS-CoV-2. A possible mechanism of this immune stimulation could be due to the fact that both Gag protein of HERV-K107 and protein S of SARS-CoV-2 share common sub-sequences as the subsequence of length 15 nucleotides from the protein S of the SARS-CoV-2 given in green on Figure 5:

GCTTTAGAACCATTT

**Homo sapiens endogenous retrovirus HERV-K102, complete sequence GenBank: AF164610.1: 1112-2596 Gag protein**

```
5'-ATGGGGCAAACTAAAAGTAAAATTAAAAGTAAATATGCCTCTTATCTCAGCTTTATTAAAATTCTTTTAA
AAAGAGGGGGAGTTAAAGTATCTACAAAAAATCTAATCAAGCTATTTCAAATAATAGAACAATTTTGCCC
ATGGTTTCCAGAACAAGGAACTTTAGATCTAAAAGATTGGAAAAGAATTGGTAAGGAACTAAAACAAGCA
GGTAGGAAGGGTAATATCATTCCACTTACAGTATGGAATGATTGGGCCATTATTAAAGCAGCTTTAGAAC
CATTTCAAACAGAAGAAGATAGCGTTTCAGTTTCTGATGCCCTTGGAAGCTGTATAATAGATTGTAATGA
AAACACAAGGAAAAAATCCCAGAAAGAAACGGAAGGTTTACATTGCGAATATGTAGCAGAGCCGGTAATG
GCTCAGTCAACGCAAATGTTGACTATAATCAATTACAGGAGGTGATATATCCTGAAACGTTAAAATTAG
AAGGAAAAGGTCCAGAATTAGTGGGGCCATCAGAGTCTAAACCACGAGGCACAAGTCATCTTCCAGCAGG
TCAGGTGCCCGTAACATTACAACCTCAAAAGCAGGTTAAAGAAAATAAGACCCAACCGCCAGTAGCCTAT
CAATACTGGCCTCCGGCTGAACTTCAGTATCGGCCACCCCCAGAAAGTCAGTATGGATATCCAGGAATGC
CCCCAGCACCACAGGGCAGGGCGCCATACCCTCAGCCGCCCACTAGGAGACTTAATCCTACGGCACCACC
TAGTAGACAGGGTAGTGAATTACATGAAATTATTGATAAATCAAGAAAGGAAGGAGATACTGAGGCATGG
CAATTCCCAGTAACGTTAGAACCGATGCCACCTGGAGAAGGAGCCCAAGAGGGAGAGCCTCCCACAGTTG
AGGCCAGATACAAGTCTTTTTCGATAAAAATGCTAAAAGATATGAAAGAGGGAGTAAAACAGTATGGACC
CAACTCCCCTTATATGAGGACATTATTAGATTCCATTGCTCATGGACATAGACTCATTCCTTATGATTGG
GAGATTCTGGCAAAATCGTCTCTCTCACCCTCTCAATTTTTACAATTTAAGACTTGGTGGATTGATGGGG
TACAAGAACAGGTCCGAAGAAATAGGGCTGCCAATCCTCCAGTTAACATAGATGCAGATCAACTATTAGG
AATAGGTCAAAATTGGAGTACTATTAGTCAACAAGCATTAATGCAAAATGAGGCCATTGAGCAAGTTAGA
GCTATCTGCCTTAGAGCCTGGGAAAAAATCCAAGACCCAGGAAGTACCTGCCCCTCATTTAATACAGTAA
GACAAGGTTCAAAAGAGCCCTATCCTGATTTTGTGGCAAGGCTCCAAGATGTTGCTCAAAAGTCAATTGC
CGATGAAAAAGCCCGTAAGGTCATAGTGGAGTTGATGGCATATGAAAACGCCAATCCTGATGTCAATCAG
CCATTAAGCCATTAA-3'
```

**Figure 5.** Complete RNA sequence of the Gag protein of the virus HERV-K102 [36]. The green subsequence of length 14 (271-285) is present in the RNA sequence of the protein S of virus SARS-CoV-2 [22]. The probability to observe this match of length 14 by chance in a sequence of 1475 nucleotides equals $10^{-6}$.

## 4. Discussion

When we combine the antibody power originated by the endogenous human retrovirus HERV-K102 envelop protein (whose part of its mRNA is shared by the SARS-CoV-2 protein S [36]) with the putative inhibitory role of circRNAs capable to block the miRNA-like action of SARS-CoV-2, one could understand why certain carriers of SARS-CoV-2 are completely asymptomatic and therefore, by mimicking their defence mechanisms, consider a possible therapy against SARS-CoV-2. Indeed, if we look on the "sponge effect" of circRNAs against microRNAs [37-39], one can consider a therapeutic effect erasing pathogenic actions of microRNAs.

For example, in the case of the human let-7e microRNA, a sub-sequence of human circular RNA PVT1 hybridizes hsa-let-7e (Figure 6), thus preventing it from exerting a too important inhibition on the translation of proteins such as the gamma-globin 2. There exists a sub-sequence of the protein S of SARS-CoV-2 (Figure 6), on which a similar action would be possible, hence reducing the miR-like pathogenicity of the protein S, but with less efficiency, with a hybridization free energy ΔG equal to -4.6 kcal/mol vs -11 for the hsa-let-7e.

**Homo sapiens Pvt1 oncogene (PVT1), long non-coding RNA NCBI Reference Sequence: NR_003367.3**

```
5'–CTCCGGGCAGAGCGCGTGTGGCGGCCGAGCACATGGGCCCGCGGGCCGGGCGGGCTCGGGGCGGCCGGGA
   CGAGGAGGGGCGACGACGAGCTGCGAGCAAAGATGTGCCCCGGGACCCCCGGCACCTTCCAGTGGATTTC
   CTTGCGGAAAGGATGTTGGCGGTCCCTGTGACCTGTGGAGACACGGCCAGATCTGCCCTCCAGCCTGATC
   TTTTGGCCAGAAGGAGATTAAAAAGATGCCCCTCAAGATGGCTGTGCCTGTCAGCTGCATGGAGCTTCGT
   TCAAGTATTTTCTGAGCCTGATGGATTTACAGTGATCTTCAGTGGTCTGGGGAATAACGCTGGTGGAACC
```
<span style="color:red">hsa-let-7e 3'–CCTTTCGATCCT</span>CCG
```
   ATGCACTGGAATGACACACGCCCGGCACATTTCAGGATACTAAAAGTGGTTTTAAGGGAGGCTGTGGCTG
   GC
```
<span style="color:red">AT</span>–5'
```
   AATGCCTCATGGATTCTTACAGCTTGGATGTCCATGGGGACGAAGGACTGCAGCTGGCTGAGAGGGTTG
   AGATCTCTGTTTACTTAGATCTCTGCCAACTTCCTTTGGGTCTCCCTATGGAATGTAAGACCCCGACTCT
   TCCTGGTGAAGCATCTGATGCACGTTCCATCCGGCGCTCAGCTGGGCTTGAGCTGACCATACTCCCTGGA
   GCCTTCTCCCGAGGTGCGCGGGTGACCTTGGCACATACAGCCATCATGATGGTACTTTAAGTGGAGGCTG
   AATCATCTCCCCTTTGAGCTGCTTGGCACGTGGCTCCCTTGGTGTTCCCCTTTTACTGCCAGGACACTGA
   GATTTGGAGAGAGTCTCACTCTGTGGTCCAGGCTGAAGTACAGTGGCATGATCCCAGGTCACTGCAACCC
```
<span style="color:green">3'–GTGAGGTATGTGAATTTTCACC–5' Protein S SARS-CoV-2</span>
```
   CCACCTCCCGGGTTCAAGTGATCCTCCTGCCTCAGCCTCCCGAGTAGCTGGTATTACAGGCGTGTGCCAC–3'
```

**Figure 6.** RNA sub-sequence of the circPVT1 [22]. The RNA sequence in red is the microRNAs hsa miR let-7 inhibited by its "sponge" hsa-circ-PVT1. The RNA sequence in green is a sub-sequence of the protein S of SARS-CoV-2 on which hsa-circ-PVT1 could serve as inhibitor. Anti-match probability of a sub-sequence of length 9 in a sequence of length 1946 is 0.06 (resp. 0.03) for the red (resp. green) sub-sequence.

We can also compare the putative miRNA-like inhibitory efficacy of the protein S in other coronaviruses than SARS-CoV-2. By taking for example the SARS CoV Rs672 virus observed in 2006, it is possible to exhibit in the RNA sequence of its protein S gene some sub-sequences similar to those from SARS-CoV-2 involved in a miRNA inhibitory effect (Figure 7): they have less nucleotides anti-matching their protein targets, which could explain lesser virulence of the SARS epidemic than of the SARS-CoV-2 outbreak.

**Bat SARS CoV Rs672/2006, complete genome GenBank: FJ588686.1: 20894-24619 protein S**

```
5'–GATTGTGTTGCTGATTACACTGTTCTCTACAACTCAA
```
<span style="color:green">CTTCATTTTCAACTTTTAAATGTTATGGAGT</span>TT
```
   CTCCCTCTAAGTTGATTGACTTGTGCTTTACAAGTGTGTATGCTGATACATTCTTGATAAGATCTTCAGA
   AGTAAGGCAAGTTGCACCAGGTGAAACTGGTGTTATTGCTGACTATAACTACAAACTGCCTGATGACTTT
   ACAGGCTGTGTCATAGCTTGGAACACTGCTAAACAAGATCAGGGCCAGTATTATTATAGATCCTCCAGAA
   AAACAAAACTTAAACCTTTTGAGAGGGATCTAACTTCTGACGAAAATGGTGTACGTACTCTTAGTACTTA
   TGACTTCTATCCTAATGTGCCTATTGAATATCAGGCTACTAGGGTTGTTGTGCTTTCATTCGAGCTTCTA
   AATGCACCTGCTACAGTTTGTGGACCTAAATTATCCACAGGACTTGTTAAGAACCAGTGTGTCAATTTCA
   ATTTTAATGGACTCAAAGGTACTGGTGTTCTGACTGATTCTTCAAAGAGATTTCAGTCATTTCAACAATT
   TGGAAGAGACACGTCGGATTTCACTGATTCCGTTCGTGACCCGCAAACATTGCAGATACTTGACATTACA
   CCATGTTCTTTTGGTGGTGTGAGTGTAATAACACCTGGAACAAATGCTTCATCTGAAGTGGCTGTTCTTT
   ACCAAGATGTAAACTGCACCGATGTCCCAACAGCCATACGTGCAGACCAATTAACACCAGCTTGGCGCGT
   TTACTCAACCGGAGTAAATGTGTTTCAAACACAAGCTGGCTGTCTTATTGGAGCTGAACATGTTAACGCT
   TCGTATGAGTGTGACATTCCTATTGGTGCTGGCATTTGTGCTAGCTAC
```
CATACAGCTTCTACTCT<span style="color:green">ACGTAGTGTAG</span>
```
   GTCAGAAATCCATTGTGGCTTACACTATGTCTTTGGGTGCAGAAAATTCTATTGCTTATGCTAA–3'
```

**Figure 7.** RNA sub-sequence of the SARS CoV Rs672 protein S gene. Nucleotides in green are homologous to those of SARS-CoV-2 protein S gene (in green on Figure 3), which could explain the lesser virulence of SARS as compared to SARS-CoV-2 due to fewer anti-matches with their miRNA-like targets. The probability to observe by chance a sub-sequence of length 31 in a sequence of 3722 nucleotides with exactly 3 errors equals $3\ C^3_{31}\ 0.25^{31} = 3\ 10^{-15}$ and for a sub-sequence of length 11 equals $9\ 10^{-4}$.

Among the symptoms of the Covid-19 disease, anosmia is frequently described. This defect could be due to a miRNA-like inhibition of mRNAs of genes from olfactory receptor family (Figure 8).

**Homo sapiens olfactory receptor family 4 subfamily E member 1 (OR4E1), mRNA NCBI Reference Sequence: NM_001317107.1**

```
5'-TTAAATGACCAAATGATTGATGCAGATACTGTGTTATATTAGACTTTTTTTCTAATTCTTTACAGGTTGT
   CTAACAAAGAGAATGGAAGAGGCCATCCTACTCAATCAAACTTCTTTAGTGACATATTTTCGGCTTAGAG
   GTTTATCTGTAAATCATAAGGCACGGATAGCTATGTTTTCCATGTTCCTCATTTTTTATGTCCTGACACT
   GATTGGGAATGTTCTCATTGTCATAACTATTATCTATGACCACCGGCTCCATACTCCCATGTATTTCTTC
   CTCAGCAACCTGTCCTTTATTGATGTCTGCCACTCCACTGTCACTGTCCCCAAGATGCTGAGAGACGTGT
   GGTCAGAGGAAAAGCTCATCTCTTTTGATGCCTGTGTGACCCAGATGTTCTTCCTGCACCTCTTTGCCTG
   CACAGAGATCTTCCTCCTCACCGTCATGGCCTATGATCGGTATGTGGCCATCTGTAAACCCCTGCAGTAC
   ATGATAGTGATGAACTGGAAGGTATGTGTGCTGCTGGCTGTGGCCCTCTGGACAGGAGGGACCATCCACT
   CCATAGCCCTCACCTCCCTTACCATCAAGCTGCCCTACTGTGGTCCTGATGAGATTGACAACTTCTTCTG
   TGATGTACCTCAGGTGATCAAGCTGGCCTGCATTGACACCCACGTCATTGAGATCCTCATTGTCTCCAAC
   AGTGGATTGATCTCCGTGGTCTGTTTTGTGGTCCTGGTGGTGTCCTACGCAGTCATCCTGGTGAGTCTGA
   GGCAGCAGATCTCCAAGGGCAAGCGGAAGGCCCTGTCCACCTGTGCAGCCCATCTCACTGTAGTTACACT
                                                            3'-ATCGATGTGA
   GTTCCTGGGACACTGCATCTTCATCTATTCCCGCCCATCCACCAGCCTCCCAGAGGACAAGGTAGTATCT
   TGCACGGGCGG-5'
   GTGTTTTTCACTGCAGTCACCCCCCTGCTGAACCCCATTATCTATACCCTTAGGAATGAAGAAATGAAGA
   GTGCCTTAAACAAGTTAGTGGGGAGAAAAGAGAGAAAAGAAGAAAAATGAAAATGTCTACGTCCTTAGGA
   TACGTGGTGCTCCAAATTAAAGAAGCGCCTTGCAAAGAATAAGTTACATACCCATAT-3'
```

**Figure 8.** Complete mRNA sequence of the human olfactory receptor family 4 subfamily E member 1 (OR4E1) [22]. The RNA sequence in green is a sub-sequence of the protein S of SARS-CoV-2, which can exert a miRNA-like inhibition of the translation of OR4E1. The probability to observe such an anti-match of length 12 by chance in a sequence of 577 nucleotides equals $5 \cdot 10^{-4}$.

## 5. Perspectives

The perspectives of the present work are in the more in-depth study of unconventional mechanisms of action of the SARS-CoV-2 virus, in particular those concerning the disturbances of oxygen transport observed in many patients [41,42]. We can also notice the resemblance of a SARS-CoV-2 sub-sequence with hsa-miR-let-7b, the microRNA the most upregulated in Kawasaki disease [43] described as potentially linked to SARS-CoV-2 infection [44]. The SARS-CoV-2 virus could have, more than a direct protein-protein interaction (proposed in [16] despite the criticisms of [45]), an effective inhibitory action *in vivo* of the same type as that predicted here *in silico* on the synthesis of subunits of human hemoglobin, and this action is more important for SARS-CoV-2 than for other coronaviruses (like the SARS CoV Rs672 on Figure 8). This hypothesis is in agreement with numerous studies showing a decrease of adult human hemoglobin blood concentrations in severe Covid-19 cases [46,47], presenting an increase of the high-sensitivity C-reactive protein as one of the three major predictors of severity [48], like in ß-thalassemia [49] and viral infections [50]. Hence, one could envisage a therapy blocking pathologic inhibitor effects on ribosomal translation of hemoglobin subunits, using for example circular RNAs as blockers of possible viral miRNA-like mechanisms (Figure 7) [51-54]. Another direction could be to search if furin cleavage site sub-sequence has the same type of interaction with key proteins like Rac small GTPase (a protein from the Rho GTPase family, which is a strong determinant of the virus-induced IFNbeta response [55-56]), implicated in replication of many important viral pathogens infecting humans or like interferons. A first example is given by the human small GTPase 1 (Figure 9) in which the inhibition of the SARS-CoV-2 protein S gene is possibly obtained through the same miRNA-like subsequence as for all type 1 interferons. The host immune system is indeed reacting to viral intrusion first with synthesis of type I interferons IFNalphas and IFNbetas [57-58]. They are messengers allowing the activation of cellular defenses blocking viral replication. In humans, these type I interferons are bound to interferon receptors, and then, they induce proteins with antiviral actions: RNA-dependent protein kinase (PKR), 2′,5′-oligoadenylate synthetase (OAS), RNase L, and Mx protein GTPases [59].

**Homo sapiens Rac family small GTPase 1 (RAC1), transcript variant Rac1, mRNA NCBI Reference Sequence: NM_006908.5**

```
5'-TAATGGAGTGAGCGTAGCAGCTCAGCTCTTTGGATCAGTCTTTGTGATTTCATAGCGAGTTTTCTGACCA
   GCTTTTGCGGAGATTTTGAACAGAACTGCTATTTCCTCTAATGAAGAATTCTGTTTAGCTGTGGGTGTGC
   TTTTTGTTACAGATTAATTTTTCCATAAAACCATTTTTTGAACCAATCAGTAATTTTAAGGTTTTGTTTG
   TTCTAAATGTAAGAGTTCAGACTCACATTCTATTAAAATTTAGCCCTAAAATGACAAGCCTTCTTAAAGC
   CTTATTTTTCAAAAGCGCCCCCCCCATTCTTGTTCAGATTAAGAGTTGCCAAAATACCTTCTGAACTACA
                              protein S SARS-CoV-2    3'-GATGT
   CTGCATTGTTGTGCCGAGAACACCGAGCACTGAACTTTGCAAAGACCTTCGTCTTTGAGAAGACGGTAGC
GATGCACGGGCG-5'
   TTCTGCAGTTAGGAGGTGCAGACACTTGCTCTCCTATGTAGTTCTCAGATGCGTAAAGCAGAACAGCCTC
   CCGAATGAAGCGTTGCCATTGAACTCACCAGTGAGTTAGCAGCACGTGTTCCCGACATAACATTGTACTG
   TAATGGAGTGAGCGTAGCAGCTCAGCTCTTTGGATCAGTCTTTGTGATTTCATAGCGAGTTTTCTGACCA
   GCTTTTGCGGAGATTTTGAACAGAACTGCTATTTCCTCTAATGAAGAATTCTGTTTAGCTGTGGGTGTGC
   CGGGTGGGGTGTGTGATCAAAGGACAAAGACAGTATTTTGACAAAATACGAAGTGGAGATTTACACTA
                              protein S SARS-CoV-2    3'-GATGTGAT
   CATTGTACAAGGAATGAAAGTGTCACGGGTAAAAACTCTAAAAGGTTAATTTCTGTCAAATGCAGTAGAT
GATGCACG-5'
   GATGAAAGAAAGGTTGGTATTATCAGGAAATGTTTTCTTAAGCTTTTCCTTTCTCTTACACCTGCCATGC
   CTCCCCAAATTGGGCATTTAATTCATCTTTAAACTGGTTGTTCTGTTAGTCGCTAACTTAGTAAGTGCTT
   TTCTTATAGAACCCCTTCTGACTGAGCAATATGCCTCCTTGTATTATAAAATCTTTCTGATAATGCATTA-3'
```

**Figure 9.** MiRNA-like subsequence of SARS-CoV-2 protein S gene (from its furin cleavage site) anti-matching a subsequence of the human GTPase 1 gene. The probability to observe such anti-matches of length 9 by chance in the of the 2301-length sequence of the whole human GTPase 1 gene equals 0.017.

**Homo sapiens interferon alpha 7 (IFNA7), mRNA NCBI Reference Sequence: NM_021057.2**

```
5'-TACCCACCTCAGGTAGCCTAGTGATATTTGCAAAATCCCAATGGCCCGGTCCTTTTCTTTACTGATGGTC
   GTGCTGGTACTCAGCTACAAATCCATCTGCTCTCTGGGCTGTGATCTGCCTCAGACCCACAGCCTGCGTA
   ATAGGAGGGCCTTGATACTCCTGGCACAAATGGGAAGAATCTCTCCTTTCTCCTGCTTGAAGGACAGACA
   TGAATTCAGATTCCCAGAGGAGGAGTTTGATGGCCACCAGTTCCAGAAGACTCAAGCCATCTCTGTCCTC
   CATGAGATGATCCAGCAGACCTTCAATCTCTTCAGCACAGAGGACTCATCTGCTGCTTGGGAACAGAGCC
   TCCTAGAAAAATTTTCCACTGAACTTTACCAGCAACTGAATGACCTGGAAGCATGTGTGATACAGGAGGT
   TGGGGTGGAAGAGACTCCCCTGATGAATGAGGACTTCATCCTGGCTGTGAGGAAATACTTCCAAAGAATC
                              hsa miR let-7b-5p  3'-CTTTGGTGTGTTG
   ACTCTTTATCTAATGGAGAAGAAATACAGCCCTTGTGCCTGGGAGGTTGTCAGAGCAGAAATCATGAGAT
GATGATGGAG-5'           3'-TGCACGGGCGGCTCCTCTTAATC-5' protein S SARS-CoV-2
   CCTTCTCTTTTTCAACAAACTTGAAAAAAGGATTAAGGAGGAAGGATTGAAAACTGGTTCATCATGGAAA
   TGATTCTCATTGACTAATGCATCATCTCACACTTTCATGAGTTCTTCCATTTCAAAGACTCACTTCTATA
   ACCACCACAAGTTGAATCAAAATTTCCAAATGTTTTCCT-3'
```

**Homo sapiens interferon regulatory factor 1 (IRF1), transcript variant 5, non-coding RNA NCBI RefSeq: NR_149069.2**

```
5'-AGAGCTCGCCACTCCTTAGTCGAGGCAAGACGTGCGCCCGAGCCCCGCCGAACCGAGGCCACCCGGAGCC
   GTGCCCAGTCCACGCCGGCCGTGCCCGGCGGCCTTAAGAACCCGGCAACCTCTGCCTTCTTCCCTCTTCC
                 3'-TGCACGGGCGGCTCCTCTTAAT-5' protein S SARS-CoV-2
   ACTCGGAGTCGCGCTCCGCGCGCCCTCACTGCAGCCCCTGCGTCGCCGGGACCCTCGCGCGCGACCGCCG
   AATCGCTCCTGCAGCAGAGCCAACATGCCCATCACTCGGATGCGCATGAGACCCTGGCTAGAGATGCAGA
   TTAATTCCAACCAAATCCCGGGGCTCATCTGGATTAATAAAGTGAGTGTAACTCTTTGGGTTTTCCTGCC
   ACTGTTTTAACCCATGTACTTCTGGAGGGACCAAAGCTTCAGATGCAGCTCAAAAAGGGAAGTGATAACG
   GGACAAGCAGGTGTTTCTCCCAGTGGGTCCTGCATGCAGGGAGTGTGCACGGCCCAGCCTGGGCCTCACT
                 3'-GTCTACGAAACTGTTATGATA-5' miRNA 301a-3p
   TGCATGACTCCTGCCTTCTTCCCTTCTTGAGGTAGGGCACCCACCTGAAGGCACTTCCAGTTTCCAGCAG
   CAAGACTTTCCAGCATCTGCAGAGCTGGAGTTCTGCTCTCCTCTAAGCGAGACCCTTACAAACATACACA
   GCACTCTGCAGGGCTCCAATCGAACAAATAGAAGACTGAGAAGTGGATGCTGCTGGGCAGAAACGTGCCT
   GGCTTAGCAGAGGACAAACGAGTTAATCTTGCACCAGTCACTCTGGCCCAAGAAGCCTATAGCTGGTGCA
   CTTGGGCAACATAGACCCTATAGACTTAGTAGCAATGATAGTATTCATAATAATAGCTAATGCTTACTG
   AACACTCCCTGTGTGCCTGGCACCTGCTAAGTATGTTATTTACATTGTGTCATTTAATCCTCGCAGTAGT
                 3'-TGATGCACGGGCGGCTCCTCTT-5' protein S SARS-CoV-2
   CCTGTGGGTTAGATCTTACTAATGTCATCATTTTCAGATAAGTAAACAGAGGCACTGAGAGGTAGATCAT
   AAGATCACACAAAAAGTGATGAAGCCAAGATTTGAACTTGAACGGTCTGACTCAGAAATCTT-3'
```

**Figure 10.** MiRNA-like subsequence of SARS-CoV-2 protein S gene (from its furin cleavage site) anti-matching sequences from the human type 1 interferon (IFNA7) or interferon regulatory factor (IRF1). In the first case, the sequence is the whole mRNA of IFNA7 and the probability to observe such an anti-match of length 8 by chance in a sequence of 730 nucleotides equals 0.04. In the second case, the sequence of the whole mRNA of IRF1 contains to targets and the probability to observe the last anti-match of length 11 by chance in a sequence of 1032 nucleotides equals $2\,10^{-3}$. In red, miRNA inhibiting sequences [59-60]. The probability to observe by chance the micro-RNA hsa miR let-7b-5p anti-match of length 9 in the first 730-length sequence equals 0.02 and the micro-RNA hsa miR 301a-3p anti-match of length 9 in the second 1032-length sequence equals 0.016.

In the same way, the miRNA-like subsequence of SARS-CoV-2 protein S gene from its furin cleavage site) anti-matches the mRNA of the MCT1 gene involved in the lactate shuttle between astrocytes and neurons (Figure 11) and this effect decreases the energy provided to the brain [61]. That could explain some neurological and neuropsychiatric complications observed in SARS-COV-2 patients, since the earliest cohorts featured non-specific neurological symptoms, such as dizziness and headache.

**Homo sapiens clone peg2135 MCT1 (MCT1) mRNA, complete cds GenBank: AY364258.1**

```
5'-ATGTTCAAGAAATTTGATGAAAAGAAAATGTGTCCAACTGCATCCAGTTGAAAACTTCAGTTATTAAGG
GTATTAAGAATCAATTGATAGAGCAATTTCCAGGTATTGAACCATGGCTTAATCAAATCATGCCTAAGAA
                    hsa miR 342-5p 3'-GAGTTAGTGTCTATCCGTGGG-5'
AGATCCTGTCAAAATAGTCCGATGCCATGAACATATAGAAATCCTTACAGTAAATGGAGAATTACTCTTT
TTTAGACAAAGAGAAGGGCCTTTTTATCCAACCCTAAGATTACTTCACAAATATCCTTTTATCCTGCCAC
ACCAGCAGGTTGATAAAGGAGCCATCAAATTTGTACTCAGTGGAGCAAATATCATGTGTCCAGGCTTAAC
TTCTCCTGGAGCTAAGCTTTACCCTGCTGCAGTAGATACCATTGTTGCTATCATGGCAGAAGGAAAACAG
CATGCTCTATGTGTTGGAGTCATGAAGATGTCTGCAGAAGACATTGAGAAAGTCAACAAAGGAATTGGCA
TTGAAAATATCCATTATTTAAATGATGGGCTGTGGCATATGAAGACATATAAATGAGCCTCAGAAGGAAT
GCACTTGGGCTAAATATGGATATTGTGCTGTATCTGTGTTTGTGTCTGTGTGTGACAGCATGAAGATAAT
               protein S SARS-CoV-2 3'-TGCACGGGCGGCTCCTCTT-5'
GCCTGTGGTTATGCT G-3'
```

**Figure 11.** MiRNA-like subsequence of SARS-CoV-2 protein S gene (from its furin cleavage site) anti-matching the mRNA of the human MCT1 gene. The probability to observe this anti-match of length 9 by chance in a sequence of 638 nucleotides equals $2.5\,10^{-3}$. In red, the micro-RNA hsa miR 342-5p inhibiting the human MCT1 gene sequence with a subsequence of length 8 and this anti-match has the probability 0.02 to occur by chance in a sequence of 638 nucleotides.

Eventually, the mutations observed on SARS-CoV-2 [35, 63-64] can be neutral (without any effect), favorable (less pathogenic) or deleterious (more pathogenic). Among them, we have (mutations in red):

| | | |
|---|---|---|
| **Neutral**: | Homo sapiens hemoglobin subunit gamma 2 (HBG2) | 5'-**GCTTTATTCTGCAAGCAA**-3' |
| | protein S SARS-CoV-2 | 3'-**TGAGGTATTGTGG**A**TTTT**-5' |
| | Homo sapiens Rac family small GTPase 1 (RAC1) | 5'-**CTGTGTGCCTGGCAC** -3' |
| | protein S SARS-CoV-2 | 3'-**GATGCACGGGCG**A**CT**-5' |
| **Favorable**: | Homo sapiens ACE2 mRNA | 5'-**GACAAATGGTCTGCCTTTTTAAAGG**-3' |
| | protein S SARS-CoV-2 | 3'-**CTT**T**TTACCG**T**CCTCGTC**AA**CAC**TT-5' |
| | Homo sapiens interferon regulatory factor 1 (IRF1) | 5'-**CCTGTGTGCCTGGCACCTGCTA** -3' |
| | protein S SARS-CoV-2 | 3'-T**GATGCACGGGCG**A**CTCCT**C**TT** -5' |
| **Deleterious**: | Homo sapiens HERV-K102 Gag protein | 5'-**GCTTTAGAACCATTT**-3' |
| | protein S SARS-CoV-2 | 5'-**G**T**TTTAGAACCATTT**-3' |
| | Homo sapiens hemoglobin betaglobin (HBB) | 5'-**TCAGAAAGTGGTGGCTGGTGTGG**-3' |
| | protein S SARS-CoV-2 | 3'-**GG**A**TTTTCACCTTTTACTACGCC**-5' |

We can notice also that the protein S gene is not the only SARS-CoV-2 gene anti-matching important human molecules. It is for example the case of the ORF10 protein with the human gamma-globin 2 (Figure 12).

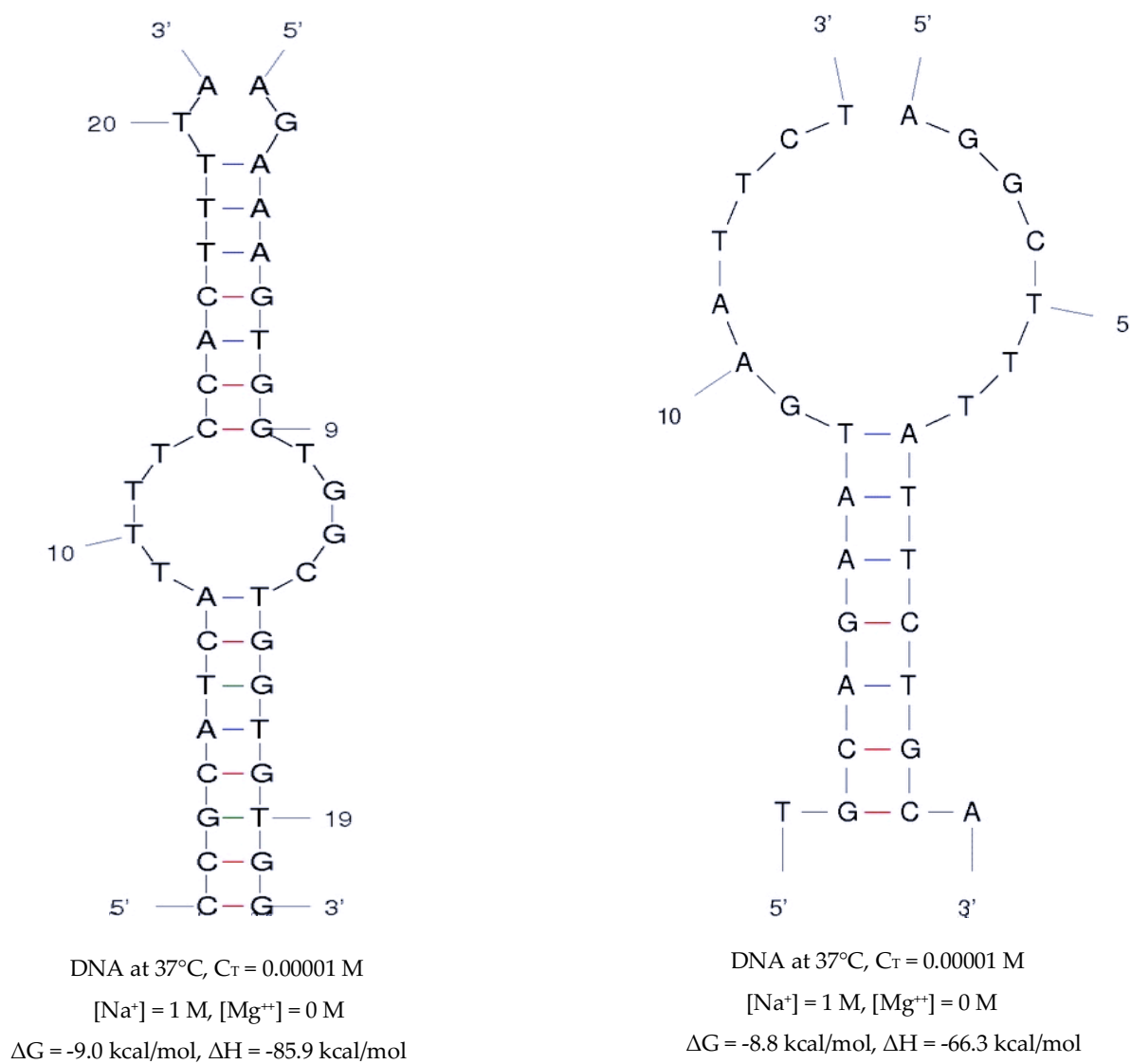

DNA at 37°C, C$_T$ = 0.00001 M
[Na$^+$] = 1 M, [Mg$^{++}$] = 0 M
ΔG = -9.0 kcal/mol, ΔH = -85.9 kcal/mol

DNA at 37°C, C$_T$ = 0.00001 M
[Na$^+$] = 1 M, [Mg$^{++}$] = 0 M
ΔG = -8.8 kcal/mol, ΔH = -66.3 kcal/mol

**Figure 12.** Hybridization between subsequences from SARS-CoV-2 genome and human genome. Left: hybridization between a subsequence of the SARS-CoV-2 Protein S gene and a subsequence of the gene of the human hemoglobin beta-globin (HBG) subunit (Figure 2). Right: hybridization between a subsequence of the SARS-CoV-2 ORF10 gene and a subsequence of the gene of the human hemoglobin gamma-globin 2 (HGG 2) subunit (Figure 1).

On Figure 12, the free energy and enthalpy are given in kcal/mol for two hybridizations [65-66] between subsequences of SARS-CoV-2 genes and subsequences of genes of two important proteins of the human metabolism of oxygen, involved in the oxygen transportation in adult for the first (the human hemoglobin beta-globin (HBG) subunit) and the in embryo for the second (the human hemoglobin gamma-globin 2 (HGG 2) subunit).

We have summarized the probabilities of anti-matches of Figures 2 to 11 and in Table 1, allowing for the comparison between the classical miRNA action and the putative inhibitory influence the protein S gene of SARS-CoV-2 can have on the translation of important human proteins.

**Table 1**. Probability P and free energy ΔG (kcal/mol) of anti-matching between human genes and protein S gene subsequence (TG and GT counting for ½).

| Matching subsequence | Human/viral gene | P | ΔG | Fig. |
|---|---|---|---|---|
| protein S **CTTCATTTTCA**ACTTTTAA**ATGTTATGGAGT** | virus SARS CoV Rs672 Protein S | $3 \cdot 10^{-15}$ | | 7 |
| protein S **GCTTTAGAACCATT** | HERV-K102 protein Gag | $10^{-6}$ | | 5 |
| **Anti-matching subsequence** | | | | |
| hsa-miR-129-5p **TGTCGTTC** | human γ-globin 2 | 0.035 | **-6.7** | 1 |
| protein S **CTGAAGTAG** | human γ-globin 2 | 0.017 | **-2.6** | 1 |
| ORF10 **AAGTAAGACGT** | human γ-globin 2 | $3 \cdot 10^{-4}$ | **-8.8** | 1 |
| RNA-dependent RNA polymerase ORF1ab **TCACGTAGA**AC**TAGGA**G**TA**TT | human beta-globin | $5 \cdot 10^{-3}$ | **-11.2** | 2 |
| miR 451a **TGAGTCAT** | human beta-globin | 0.04 | **-3.2** | 2 |
| protein S **TTTTCACC** | human beta-globin | 0.017 | **-9** | 2 |
| miR 92a-1-5p **TGTCCGGC** | human beta-globin | 0.04 | **-8.2** | 2 |
| protein S **CTGTTTACCG** | human ACE2 | $3 \cdot 10^{-3}$ | **-9.5** | 4 |
| protein S **GTGAGGTAT** | human circPVT1 | 0.03 | **-4.6** | 6 |
| let-7e **CCTTTCGAT** | human circPVT1 | 0.06 | **-11** | 6 |
| protein S **ATCGATGTGATG** | human olfactory receptor OR4E1 | $5 \cdot 10^{-4}$ | **-7.7** | 8 |
| protein S **GATGTGATG** | human GTPase 1 | 0.017 | **-6.8** | 9 |
| let-7b-5p **CTTTGGTGT** | human type 1 interferon IFNA7 | 0.04 | **-3.2** | 10 |
| protein S **GCACGGGC** | human type 1 interferon IFNA7 | 0.04 | **-10.7** | 10 |
| miR 301a-3p **GTCTACGAA** | human type 1 interferon IRF1 | 0.016 | **-3.6** | 10 |
| protein S **GATGCACGGGC** | human interferon regulatory factor IRF1 | $2 \cdot 10^{-3}$ | **-8** | 10 |
| miR 342-5p **GAGTTAGT** | human MCT1 | 0.02 | **-3.9** | 11 |
| protein S **GCACGGGCG** | human MCT1 | $2.5 \cdot 10^{-3}$ | **-4** | 11 |

## 6. Conclusion

To conclude, the natural history of the SARS-CoV-2 virus remains widely unknown and it is still too early to say whether the many mutations observed will cause it to evolve in a favorable direction from a human point of view. There are for example some mutations surely deleterious [66,67], but also others favoring the positive role of some human miRNAs against SARS-CoV-2 [68-70] suggesting a possible therapy. The present proposal of a miRNA-like mechanism would at least allow to see, for a predictive purpose, what mutations are keeping, losing or reinforcing its pathogenicity.

**Author Contributions:** Conceptualization, methodology, investigation, J.D.; resources, J.D.; data curation, J.D.; writing—original draft preparation, J.D.; writing—review and editing, H.S. All authors have read and agreed to the published version of the manuscript.

**Funding:** This research received no external funding

**Conflicts of Interest:** The authors declare no conflict of interest.